# Advances in the Physics of Magnetic Skyrmions and Perspective for Technology

Albert Fert, Nicolas Reyren and Vincent Cros

*Unité Mixte de Physique CNRS, Thales, Univ. Paris-Sud, Université Paris-Saclay, 91767 Palaiseau, France*

**Abstract**

Magnetic skyrmions are small swirling topological defects in the magnetization texture stabilized by the protection due to their topology. In most cases they are induced by chiral interactions between atomic spins existing in non-centrosymmetric magnetic compounds or in thin films in which inversion symmetry is broken by the presence of an interface. The skyrmions can be extremely small with diameters in the nanometer range and, importantly, they behave as particles that can be moved, created or annihilated, making them suitable for "abacus"-type applications in information storage, logic or neuro-inspired technologies. Up to the last years skyrmions were observed only at low temperature (and in most cases under large applied fields) but important efforts of research has been recently devoted to find thin film and multilayered structures in which skyrmions are stabilized above room temperature and manipulated by current. This article focuses on these recent advances on the route to devices prototypes.

**Introduction**

Schematic representations of the magnetization configuration in a magnetic skyrmion standing in a thin film uniformly magnetized in the up direction outside the skyrmion are displayed in Figures 1 a-b. The magnetization inside the skyrmion rotates progressively with a fixed chirality from the up direction at one edge to down at the center and again up on the opposite edge. Fig. 1a (Néel-like skyrmion) and Fig. 1b (Bloch skyrmion) correspond to different symmetries of the structures (due to the underlying crystal lattice or the presence of an interface, for instance) and the resulting different directions of the rotation. In most systems the spin configuration of skyrmions is caused by chiral interactions of the Dzyaloshinskii-Moriya type[1,2]. Even though other types of localized magnetic texture, e.g. magnetic bubbles, can be stabilized without Dzyaloshinskii-Moriya Interactions (DMI), e.g. by dipolar interaction, a crucial difference is that they would not have the defined chirality that impacts most of the properties of the skyrmions[3,4,5]. Skyrmions can in fact be defined by the topological number *S* (or skyrmion number) characterising the winding of the normalized local magnetization, ***m***, which is, in the two-dimensional limit:

$$S = \frac{1}{4\pi} \int \boldsymbol{m} \cdot (\partial_x \boldsymbol{m} \times \partial_y \boldsymbol{m}) \mathrm{d}x \mathrm{d}y = \pm 1, \qquad \text{(Eq. 1)}$$

The normalized magnetization can be mapped on a unit sphere and, in the case of skyrmions, it covers the entirety (4π) of the sphere. This peculiar topology governs the skyrmion properties[6]. In particular, it gives rise to a topological protection of the structure: the spin configuration cannot be twisted continuously to obtain a configuration of different *S*, a ferromagnetic one for example; there is a topological barrier stabilizing the skyrmion[7]. Finally, a crucial property of magnetic skyrmion is their solitonic nature: its finite extension allows it to move as a particle and/or to be excited at specific modes. The dynamics of these objects can hence be used for interesting devices that will be described in more details hereafter.

Magnetic skyrmions were identified initially in single crystals of magnetic compounds having a non-centrosymmetric lattice[8,9] and explained by the existence of DMIs induced by spin-orbit coupling in the absence of inversion symmetry in the lattice of the crystal. Skyrmions were then observed in ultra-thin magnetic films epitaxially grown on heavy metals and submitted to large DMIs induced by the breaking of inversion symmetry at the interfaces together with the strong spin-orbit coupling of the heavy metal. The first systems were monolayers of Fe and bilayers Fe/Pd on heavy metal substrates on Ir(111), where the DMI can be quite large at the Fe/Ir(111) interfaces[10]. As shown in Figs. 1c, these skyrmions are extremely small, extending over a few lattice parameters only, but the stabilization of the skyrmion phase necessitates large magnetic field (of the order of 1 T) and low temperatures (below about 30 K for a Fe monolayer). Moreover, in systems such as one monolayer of Fe on Ir(111), a skyrmion lattice is the ground state, single skyrmions cannot exist and therefore the properties of individual particle cannot be really exploited. For a Fe/Pd bilayer grown epitaxially on Ir(111), a spin spiral is observed at low field but an applied field induces a transition to a ferromagnetic state embedding individual metastable skyrmions[10,11], as in Fig. 1d for an applied field of about 1T. The conditions for the having individual skyrmions instead of periodic spin textures e.g. skyrmion lattices or spin spirals, are discussed in BOX-1.

For applications of skyrmions, a prerequisite is the stabilization of small individual skyrmions at room temperature (RT) and in zero or very small applied field. As, up to now, the Néel or Curie temperatures of bulk compounds in which skyrmions have been found are generally below or just around RT, this way is not really attractive as long as skyrmions have not been found in non-centrosymmetric compounds with higher ordering temperature. Although thin films of ferromagnetic transition metals as Fe or Co are more tempting, ultra-thin epitaxially grown films do not seem the most convenient way for devices, first because, up to now, skyrmions in these ultra-thin films have been found only at low T and second because epitaxial growth is not easily compatible with the usual spintronic technologies. A prominent direction toward practical RT individual skyrmions has been the recent development of perpendicularly magnetized multilayers prepared by sputtering deposition and exploiting the possibility of additive DMI at successive interfaces. Successful results have been recently obtained in several groups, not only on the stabilization of skyrmions at room temperature but also on their current-induced manipulation, creation and displacement. Our article will focus on the recent results in this new field of topological spintronics in which topology together with chiral interactions and spin-orbit torques are exploited in an entirely new context for application in the future generation of the information and communication technologies.

**Interfacial DM interaction**

In systems lacking inversion symmetry, spin-orbit coupling (SOC) can induce a chiral Dzyaloshinskii-Moriya Interaction (DMI), which takes the form of:

$$H_{DMI} = -(\mathbf{S}_1 \times \mathbf{S}_2) \cdot \mathbf{d}_{12} \qquad (Eq.2)$$

Here $\mathbf{S}_1$ and $\mathbf{S}_2$ are neighbouring spins and $\mathbf{d}_{12}$ is the corresponding DM vector. The DMI is a chiral interaction lowering or increasing the energy of the spins depending on whether the rotation from $\mathbf{S}_1$ to $\mathbf{S}_2$ around $\mathbf{d}_{12}$ is clockwise or counter-clockwise. If $\mathbf{S}_1$ and $\mathbf{S}_2$ are initially parallel, the effect of a strong enough DMI (compared to exchange interaction) is to introduce a relative tilt around $\mathbf{d}_{12}$. In magnetic films with interfacial DMI, the DM vector lies in the plane of the film (x-y plane) and the global effect of DMI on it magnetization $\mathbf{m}$ can be expressed by the micromagnetic energy per volume

$$E = D \cdot (m_z \partial_x m_x - m_x \partial_x m_z + m_z \partial_y m_y - m_y \partial_y m_z) \qquad (Eq.3)$$

where $D$ is related to the pair interaction $d_{12}$ of Eq.(2) and, for purely interfacial DMI, inversely proportional to the thickness of the film ($D$ is positive for anticlockwise rotation).

The existence of DMI has been first proposed to account for the properties of magnetic compounds having a non-centrosymmetric lattice such as α-$Fe_2O_3$[1,2]. The existence of DMI was theoretically understood as an additional term induced by spin-orbit coupling in the super-exchange interaction of magnetic insulators in the absence of inversion symmetry (Moriya). For metallic systems, the existence of a chiral interaction have been first demonstrated for disordered alloys in which an atom with large SOC mediates a DMI between two magnetic atoms with $\mathbf{d}_{12}$ perpendicular to the plane of the triangle formed by the three atoms[12]. Then DMI was also predicted at the interface between magnetic films and metal of large SOC[13]. In the simplified picture of Fig. 2a for an interface between a magnetic film (here Co) and a strong SOC metal (here Pt), the DMI can be seen as induced by Co-Pt-Co triangles with a vector $\mathbf{d}_{12}$ in the plane of the interface and perpendicular to the vector connecting $\mathbf{S}_1$ and $\mathbf{S}_2$. Experimentally spin textures induced by interfacial DMI have been first observed in epitaxial ultra-thin magnetic films, first spin spirals for 2 monolayers (ML) of Fe on W(110)[14], even though the origin of the effect was not clear yet. Then similar texture were identified as interfaced-induced DMI for Mn on W(110)[15,16] for example, and later with nanoscale skyrmion lattices on 1 ML Fe grown on Ir(111)[10]. DMI at various types of interfaces between magnetic films and heavy metals such as Pt, Ir or W have been now calculated consistently by *ab-initio* studies, for example $d_{12}$ = 1.7 eV for 1 ML of Fe on Ir(111) surface and -1.8 eV for 1 ML of Co on Pt(111) surface by Dupé *et al*[17], or $d_{12}$ = -1.9 eV for 1 ML-Fe/3 ML-Pt(111) and $d_{12}$ = -1.7 eV for 1 ML-Co/3 ML-Pt(111)[18]. Series of results from Belabbes *et al*[19] are displayed in Fig. 2b. These first principle calculations are useful to provide a deeper insight in the physics of the interfacial DMI and also a guideline for materials and interface engineering. As displayed in Fig. 2c from Yang *et al.*[18] for an interface between Co and Pt, the DMI is quasi-purely interfacial, i.e., much stronger in the Co atomic layer just at the interface and almost negligible in the other atomic layers. As shown by Belabbes *et al.*[19], the relative position of the $3d$ states of the magnetic transition metal and $5d$ states of the heavy metal is an important parameter controlling their hybridization at the interface, the resulting charge transfer between $3d$ and $5d$, and the interfacial electric field. The DMI is smaller when the heavy metal $5d$ levels are shifted below the magnetic metal $3d$ levels as for interfaces with Au.

Finally, it has been found in recent experiments that DMI can also be produced not only at the interfaces with $5d$ metals but also with oxides as MgO[20] or also graphene[21]. These DMIs cannot be described in the picture for purely metallic interfaces but, according to density functional theory (DFT) calculations, are related to a particularly large charge transfer and interfacial electric field that compensates the small SOC of the atoms at the interface[20,21]. For similar reasons, interface oxidation is expected to increase DMI[22].

Several types of experiments have been used to measure directly the DMI at different types of interfaces such as Brillouin Light Scattering[23,24,25], propagation[26,27] or nucleation[28] of chiral magnetic domain walls (DW). Note that a precise quantitative comparison of the DMI estimation from these different techniques for a given multilayer system has not been systematically done[29]. Among these approaches, Brillouin Light Scattering, at least for systems made of a single magnetic thin film, is the one that requires fewer material parameters usually extracted from other experiments. However, what is important for the skyrmion stabilization is that very large DMI were found in several systems. Their interfacial character has been demonstrated by the variation of DMI with the film thickness[30].

For the formation of skyrmions in magnetic films with out-of-plane magnetization, the DMI is in competition with the exchange interaction (characterized by the stiffness constant *A*) that tends to align the spin and the out-of-plane anisotropy (characterized by the anisotropy coefficient *K*) that tends to keep the spin in the perpendicular orientation. As described in BOX 1, the stabilization of individual skyrmions is mainly controlled by the interplay between DMI, exchange and anisotropy. Additional effects come from magnetostatic interactions and the geometrical confinement of the structure.

**Individual skyrmions at room temperature in bilayers and multilayers**

Series of promising results at RT have been recently obtained in multilayers combining interface-driven out-of-plane magnetic anisotropy with additive DMI at successive interfaces[31,32,33,34]. In systems such as the (Ir1nm/Co0.6nm/Pt1nm)x10 multilayers of Fig. 3a, the additive DMI at the Ir/Co and Co/Pt interfaces[17,18] induce skyrmions in the Co layers. As the skyrmions in successive Co layers are coupled through the ultra-thin nonmagnetic layers, the larger magnetic volume of the resulting columns of coupled skyrmions can stabilize them at RT even if the diameter of the column is small. Importantly this type of multilayers can be fabricated by standard technologies of spintronic devices, generally by magnetron sputtering.

In the mapping (Fig. 3b) of the out-of-plane magnetization from a (Ir/Co/Pt)x10 multilayer derived from dichroic signals in Scanning Transmission X-ray Microscopy (STXM), a disordered ensemble of individual skyrmions can be identified in the images of the 10-70 mT field range between a phase with maze magnetic domains at lower field and ferromagnetic saturation above 80 mT[31]. The size of the skyrmions can be derived from a fit of their magnetization profile with a standard profile for compact skyrmion taking in account the resolution of the instrument (see Fig. 3 and caption). The field dependence of the radius of a given skyrmion, varying between 15 and 40 nm in Fig. 3c, is consistent with the result of micromagnetic simulations of individual skyrmion with a DMI coefficient $D$ around 1.9 mJ/m$^2$ (Moreau-Luchaire et al[31]). The skyrmions in (Ir/Fe/Co/Pt)x20 multilayers shown in the MFM images of Fig. 3d are also obtained in an intermediate field range between zero field and ferromagnetic saturation[32]. The evolution from disordered individual skyrmions and skyrmion lattice as a function of the relative thicknesses of Co and Fe is displayed in Fig. 3d (see details and discussion in BOX-1). Skyrmions at RT have also been obtained in several other types of multi-layered structures, Ta/CoFeB/TaO$_x$ trilayers[35] (radius ~1 µm), multilayers such as (Pt/Co/Ta)x15 (radius ~120 nm) and (Pt/CoFeB/MgO)x15 (radius ~120 nm) multilayers[33], Pt/Co/MgO trilayers[20] (Fig. 3e, radius ~65 nm), and Ta/Co/MgO (radius ~450 nm) with an inserted Ta layer of variable thickness between Co and MgO[34]. The skyrmions in Fe/Ni bilayers[36] (radius ~100 nm) displayed in Fig. 3f are stable at RT and also in zero field, thanks to a field-like interaction induced by exchange coupling with a Ni underlayer through a Cu spacer layer (Fig. 3g).

All the results at RT described above are promising for rapid advances on the route to skyrmionic devices. However, there are still several pending questions to clear up. First, there is a very large dispersion of the skyrmion sizes in different observations. Diameters below 100 nm have been obtained only by Moreau-Luchaire et al.[31] and Soumyanarayanan et al.[32], down to 30 nm (under field of a few tens of mT), relatively close to what can be required for most applied memory devices (< 10 nm). In many results, the size is in the 100 nm-1 μm range and in other results, sizes above 1 μm[35] probably indicate that the spin configuration does not correspond to a compact skyrmion but rather to a chiral bubble (as discussed in Box 2, chiral or skyrmionic bubbles, domains with inversed magnetization surrounded by a Néel wall with the chirality imposed by DMI, have a topological number generally equal to unity and share many properties with compact skyrmions). Actually the dispersion of the observed sizes can be understood if one realizes that the size depends on the respective values of many parameters, DMI constant $D$, exchange stiffness $A$, out-of plane anisotropy $K$ and dipolar fields. Importantly, when $D$ is close to the critical $D_c$ corresponding to zero energy domain wall, the effect of the dipole fields becomes particularly important[20] and tends to increase significantly the skyrmion size. Structures with reduced dipole fields and appropriate combinations of the different parameters can probably be found to decrease the size below 10 nm as desired for memory applications.

Another issue for skyrmion-based devices is the possibility to stabilize skyrmion without external applied field. This can be done by confinement effects or by using a local exchange or dipole field (see Fig. 3g) generated by an additional magnetic element[36,37,38].

**Current-induced motion of skyrmions**

Magnetic skyrmions will be inserted in devices only if they can be moved and/or excited at low energy costs. The motion of skyrmion lattice induced by a spin polarized current was already observed in very early experiments. First, in 2010, Jonietz et al.[39] detected a current-induced rotation of the skyrmion lattice in MnSi and, in 2012, a direct observation of current-induced motion was obtained by Yu et al.[40]. The effect of electrical current can be described as a result of the spin transfer torque (STT) exerted by spin-polarized currents (or pure spin currents) on the magnetization[41,42] or, alternatively, as a consequence of the emergent electromagnetic fields generated by the skyrmion spin texture[43,44]. A general feature of the motion is the co-existence of a longitudinal motion (along the current) and a transverse one generated by gyrotropic effects related to the topology of the skyrmion.

Spin torques can be produced either by the spin-polarized current flowing inside the ferromagnetic film (CIP-STT in Fig. 4a) or by a vertically injected spin current, for example the one generated by the Spin Hall Effect (SHE-STT in Fig. 4a) in the heavy metal layer (SHE injects a vertical pure spin current of density typically equal to the horizontal charge current density in the heavy metal multiplied by the SHE angle). In Fig. 4a, examples of results of micromagnetic simulations[41] of the skyrmion velocities in a Co nanoscale track for both injection geometries are shown. For simplicity, only the results for SHE (larger velocities are expected for the same current density with SHE-STT) will be described. As shown in Fig. 4b, while moving the skyrmion has both longitudinal and transverse velocities during an initial transient regime (< 1 ns in these simulations) but the transverse motion, often dubbed as skyrmion Hall effect, is stopped at some distance from the edge due to the repulsive interaction from the edges. The final steady motion is parallel to the track (and so parallel to the current) and its variation with current is displayed in Fig. 4a. Velocities as large as 10 m/s are obtained for a horizontal current density of 4 MA/cm$^2$ in a heavy metal of spin Hall angle of 0.1 (the simulation takes only into account the predominant anti-damping torque). Note that the skyrmion velocity is expected to be inversely proportional to the damping coefficient α, thus offering substantial room for improvement as many magnetic materials show much smaller Gilbert damping coefficient than $\alpha = 0.3$ (Co) used in the simulations of Fig. 4a. The direction of the longitudinal motion depends on both the chirality of the skyrmion and the sign of the spin Hall angle in the heavy metal. Typically, the motion is against the electron flow for the chirality of skyrmions induced by Co/Pt interfaces and the sign of the spin Hall angle in Pt. The deflection of the skyrmion in transversal directions, left or right, is depending on the up or down polarization at the centre of the skyrmion. It can also be noticed that, at large enough current density, the skyrmion will eventually reach the edge and be expelled or slides along it, what corresponds to the end of the SHE-STT curves in Fig. 4a.

The current-induced motion of the skyrmions can be analytically described from the Thiele equation[45]

$$\mathbf{F} + \mathbf{G} \times \mathbf{v} + \alpha \mathcal{D}\mathbf{v} = 0 \tag{Eq.4}$$

where **F** is the force exerted by SHE (**F**$_{SHE}$) and repulsion by the track edges, **G** is the gyrovector oriented along the out-of-plane direction and proportional to the skyrmion number, **v** the velocity, α the damping coefficient and $\mathcal{D}$ the dissipative tensor that can be calculated from the parameters of the skyrmion[41,42,45,46,47,48]. For a skyrmion in an infinite medium and, for simplicity, neglecting the term due to the Oersted field, the SHE-induced longitudinal/transverse velocities as well as the skyrmion spin Hall angle *Θ* defined as the angle between the motion and the current direction and characterizing the so-called skyrmion Hall effect can be expressed as[48]:

$$v_x = \frac{\delta}{1+\delta^2} \frac{F_{SHE}}{G_z} \tag{Eq.5}$$

$$v_y = \frac{1}{1+\delta^2} \frac{F_{SHE}}{G_z} \tag{Eq.6}$$

$$\theta_{SH} = \mathrm{atan}\left(\left|\frac{v_y}{v_x}\right|\right) = \mathrm{atan}(1/\delta) \tag{Eq.7}$$

with $\delta = \alpha \mathcal{D}_{xx}/G_z$, where $\mathcal{D}_{xx}$ is the diagonal element of the dissipative tensor and $G_z$ the out-of-plane component of the gyrovector. When the skyrmion is in a track, the repulsive force by the edge leads to the following expression of the final longitudinal velocity $v_x^*$ along the edge[41,48]

$$v_x^* = \frac{1}{\delta} \frac{F_{SHE}}{G_z} \qquad (Eq.8)$$

The main features of the motion that can be derived from the above equations and the dependence of $\delta$ and $F_{SHE}$ on the parameters $\alpha$, $A$, $K$ and skyrmion radius $R$ defined as the radius of the circle with zero out-of-plane magnetization. Taking the approximation standing for large radii compared to[48] $\Delta$ (where $\Delta = (A/K)^{1/2}$ is the length scale of a DW width), $\delta \approx \alpha(R/2\Delta)$ and $\frac{F_{SHE}}{G_z} \propto R$, which leads to the dependence of the longitudinal/transverse velocities (Fig. 4d) and the skyrmion Hall angle (Fig. 4c) as a function of the size and the damping coefficient $\alpha$. Interestingly one also finds from Eq. (8) that the repulsion by the edges of a track leads to a velocity $v_x^*$ definitely larger than the velocity $v_x$ in free space and, at least in the approximation for large skyrmionic bubbles, independent of the skyrmion size. (See also Iwasaki *et al.*[49])

The demonstration of current-induced motion of skyrmions in several types of multilayers followed shortly the first observations of skyrmions at room temperature in magnetic multilayers in 2015-2016. The MOKE microscopy images of Ta/CoFeB-1.1nm/TaO$_x$ trilayers[35] shown in Fig. 5a give an example of successive displacements of skyrmions (actually chiral bubbles) induced by current pulses (with motion ascribed to the SHE in Ta). One sees clearly the skyrmion Hall angle between the motion and the current. Similar observations of skyrmion Hall effect have been obtained in (Pt/CoFeB-0.7nm/MgO)x15 multilayers by Litzius *et al.*[50]

In both experiments by Jiang *et al.*[35] and Litzius *et al.*[50], the directions of the longitudinal and transverse motions are in agreement with the sign of the SHE and chirality of the skyrmions. What has not been clearly experimentally established yet, is a quantitative understanding of the motion with definite results about the skyrmion (or chiral bubble) velocity dependence on the size, current density and concentration of defects. When we look at the variation of the velocity along the motion direction[35] with the current density in the inset of Fig. 5a, for about 1 µm and also for 200 nm skyrmionic bubbles in (Pt/CoFeB-0.7nm/MgO)x15 multilayers[33] shown in Fig. 5b, a common observation is that the motion starts only above a critical current density, which can be attributed to pinning by defects and suggests a creep-like motion around the critical current. Above the critical region, the velocity in Fig. 5b and 5c increases almost linearly and reaches very large values around 100 m/s at current densities around 5x10$^{11}$ A/m². For the bigger skyrmions (about 1 µm) of Fig. 5a[35], the velocity also increases almost linearly with the current but the slope is around 10 times smaller than in Fig.5b. One could be tempted to look back to Fig. 4 and ascribe the velocity difference to the different diameters, about 200 nm in (Pt/CoFeB-0.7nm/MgO) in Fig. 5b and 10 times larger for the skyrmionic bubbles in Fig. 5a. However it also turns out that small velocities have also been found for skyrmions in the 100 nm range[51], so that the understanding on the role of the size on the velocity is not clearly established yet. In addition, in both experiments by Jiang *et al.*[35] and Litzius *et al.*[50], the skyrmion Hall angle increases with velocity, which is also inconsistent with the simple model based on Thiele equation (Eq. 7) for skyrmions or chiral bubbles in films without defects[52].

Several other interesting issues have not been really tackled yet in the recent publications described above. First most experiments have been performed on skyrmions with sizes above 100 nm even though skyrmions in the 10 nm range are the most interesting for applications. Second the increase of the velocity along edges by the repulsion force for skyrmions in a track, see discussion of Eq. 8, is still a pending question, as well as the related problem of the motion of very small skyrmions in narrow tracks. We have as well to understand better what sample homogeneity will be required to obtain uniform motions for applied devices. On the fundamental side also, there are not yet clear experimental results on the question of the skyrmions inertia[53].

The current-induced motion of skyrmions has also been investigated in more complex situations as, for example, for two perpendicularly magnetized ferromagnetic layers strongly coupled by an antiferromagnetic interactions[54,55,56]. When a skyrmion is created in one of the layers, it generates a skyrmion of the same chirality but opposite polarity in the second layer. Their current-induced longitudinal motions are in the same direction but, for strongly coupled skyrmions, there is compensation between their opposite transverse deflection and the motion is purely longitudinal[54,55,56]. In another type of structure, Hrabec *et al.*[48] proposed to use the dipolar coupling between ferromagnetic layers (actually not single layers but Ni/Co/Ni trilayers) separated by an Au layer and interfaced with top and bottom Pt layers in order to obtain a state with ferromagnetically coupled skyrmions of opposite chiralities in the two FM stacks. These skyrmions can be moved at high velocity by the SHE-STT from the Pt layers, as shown in Fig. 5c, with a similar dependence on the current as in Fig. 5a-b.

Note that skyrmion motion could also be controlled using magnons or temperature gradients[57,58].

Finally, the main conclusion is that, after one year of experiments on RT current-induced motion in multilayers, we know already that skyrmions can be moved at relatively high speed with currents comparable to those needed for similar speeds in the motion of Néel DW induced by DMI[59,60,61,62]. These first results are not fully understood yet but are already promising towards the targeted applications. On the route to devices, it will be necessary to extend the current-induced motion to smaller compact skyrmions and also to improve the homogeneity of the structures for applications requiring a coordinated motion of individual skyrmions.

*Nucleation and detection of skyrmions*

As we will see in the last section, the combination of their topological nature and their chiral properties imposed by the DMI makes the magnetic skyrmions very promising for several classes of future spintronic devices. However, whatever the targeted applications using skyrmions, an efficient writing and reading process will have to be implemented. Different strategies have been already proposed for achieving a reproducible nucleation of individual skyrmions under request and at a given position of a device. Indeed, several concepts have been first investigated through numerical simulations of current-induced effects. For example, Iwasaki *et al.*[42] proposed a creation process of skyrmions that relies on CIP-spin transfer torques deflected by an artificial notch (see Fig. 6a). A way to avoid the introduction of a local defect which might be a problem for the device scalability is to inject the spin polarized current in a perpendicular geometry as shown in Fig. 6b. In such a geometry, Sampaio *et al.*[41] showed that the dynamical reversal of the magnetization due to spin torques under the injection contact at the centre of a Pt/Co nanodot can lead to the formation of a magnetic skyrmion. In fact, the nucleation and also the deletion of individual skyrmions by the injection of a localized spin polarized current has been observed experimentally by Romming *et al.*[11] using the injection through the tip of a STM in the FePd bilayer system. Another original nucleation process has been developed by Jiang *et al.*[35] (Fig. 6c): it relies on inhomogeneity of the current lines at the exit of the constriction that generate a multidirectional spin torque[63] acting on the different parts of a DW (see Fig. 6d) and thus leading to the formation of micron-size magnetic skyrmionic bubbles[64]. Note that this efficient nucleation scheme has not been yet experimentally demonstrated for small compact skyrmions. A similar approach based on the conversion of a domain wall in a conduit into a skyrmion has been studied numerically by Zhou *et al.*[65] and is at the basis of several conceptual skyrmion based devices for logic applications as we will see later. Laser-induced nucleation of magnetization texture is yet another possibility[66]. Finally, Hsu *et al.*[67] have established the possibility to use a local electric field (from a non-magnetic tip) instead of spin polarized current to switch reversibly the magnetic configuration of triple layer of Fe deposited on Ir(111) between a uniform magnetized state and a skyrmion state (see Fig. 6e). This successful observation of nucleation and deletion through this low energy process might represent a key breakthrough in the case that it can be demonstrated at room temperature in multi-layered skyrmion systems.

In view of future spintronic applications based on magnetic skyrmions, another key issue is the prospect of electrically detecting the presence of individual skyrmions. In fact, several magneto-transport phenomena can be envisaged. A relatively standard approach along with what has been already proposed for domain wall race-track memory will be to rely on the magnetoresistive response induced by the passage of a skyrmion under a magnetic tunnel junction (MTJ) located on top of a racetrack (see Fig. 7a). Note that taking advantage of the topological protection associated to skyrmions, it would also be possible to design a multi-level memory using the number of skyrmions to code different states, each MTJ element being readable by the tunnel resistance which depends on this number. Another approach for efficient detection could be to take advantage of the non-collinear magnetoresistance (NCMR) arising due to changes in the band structure[68] associated with the large magnetization gradient in the presence of a skyrmion (Fig.7b)[69]. Note however that this NCMR effect does not depend intrinsically on the special topological properties of the skyrmion but rather on the fact that skyrmions present a highly non collinear spin texture. A similar comment can be made if one considers to use the longitudinal anisotropic magnetoresistance (AMR) or the anomalous Hall effect (AHE)[70] for detecting the presence of individual skyrmions. Indeed, both these effects will generate a signal that is only proportional to the *z*-component of the magnetization. More originally and directly related to the topological properties of skyrmions, a possible scheme for skyrmion detection is to leverage on the so-called topological Hall effect (THE)[6]. This effect arises from the accumulated Berry phase acquired by conduction electrons (in the adiabatic limit) when they are passing through a skyrmion. The Berry phase is proportional to the skyrmion number (and not to the amplitude of the spin-orbit interaction) and corresponds to an emergent magnetic field produced by the skyrmion and acting on the electrons. This emergent field hence results into a transverse deflection of the electrons and thus leads to a measurable Hall voltage allowing a purely electrical detection of skyrmions[71] as shown in Fig. 7c and d. It is to be noticed that the existence of a large topological Hall voltage has been observed in bulk non centro-symmetric materials[72,73] just after the discovery in 2009 of the skyrmion lattice phase existing in these materials. More recently, by measuring at low temperature discrete changes of the topological Hall resistivity of nanoscale FeGe Hall-bar structures (see Fig. 7d), Kanazawa *et al.*[74] not only demonstrated for the first time the quantized nature of the emergent field in each skyrmion but also proved that a purely electrical detection of single nanoscale skyrmion is feasible. Up to our knowledge, there is not yet a clear experimental evidence of a large THE transverse voltage detected at room temperature existing in metallic multilayer containing skyrmions, in which the THE signal is expected to be much smaller.

**Perspective for 'topological' room temperature applications**

Magnetic skyrmions might offer a unique opportunity to bring topology into room temperature electronic devices for information and communications technology (ICT). Even though topological properties are ubiquitous in some of the most exciting recent advances in condensed matter physics, e.g. high-temperature interfacial superconductivity, quantum Hall effect, or topological insulators, magnetic skyrmions hold a special place as they are probably the most promising candidates to be used at medium term into consumer low energy nanoscale spintronic devices.

The new field of research leveraging on skyrmion topological properties is just arising and a lot of issues remain to be tackled as described in previous sections. However, all the basic functions, i.e., writing the information (nucleation of individual skyrmions), processing the information (displacement, creation/annihilation of skyrmions, excitation of skyrmion modes), reading the information (electrical detection of individual skyrmions) have been already demonstrated separately but often only at low temperature and for skyrmion lattice rather than for single skyrmion at RT. The next big challenge will be to achieve these three functionalities at room temperature and using all electrical schemes in single compact integrated devices. Hereafter we shortly review examples of what we believe are some of the most interesting skyrmionic device concepts recently proposed (for an extensive review of application of skyrmions, we refer to W. Kang *et al.*[75]).

*Skyrmion race track memory*

The first one is the so-called skyrmion racetrack memory whose principle is very similar to the one based on DWs proposed by S. S. P. Parkin[76]. Exploiting the solitonic character of skyrmions, the information can be coded by a sequence of individual skyrmions in a magnetic track[41,46,77] (Fig.8a). One of the advantages of skyrmions over domain walls is their high level of integration[78]. As shown in Fig. 8b, the calculated diameter of a 'free' compact skyrmion can be compressed by a few folds by decreasing the track width and, in current-induced motion, the skyrmion follows precisely the middle line of the track because of this confinement. Moreover, as it has been also shown though simulations[41,79] that the spacing between neighbouring skyrmions in a track can be of the order of the skyrmion diameter, one can expect a higher density with skyrmions than with DW in a racetrack memory. As for the energy consumption, even if the ratio of velocity over current density $v/J$ is not expected to be very different for DW and skyrmions, the flexibility of their shape and/or their trajectories should allow the skyrmions to move with extremely small depinning currents, as it has been observed in skyrmion lattices in bulk materials[39,40]. Another advantage of skyrmions is that their motion by spin torques will be similar in straight tracks or in curved ones as they are guided by the confinement from the edges, whereas the motion of DWs will be affected in curved parts of the race track because the torques will act differently in the wall at the inner and the outer parts of the track. At last, the passing of skyrmions in the track can be counted through standard TMR devices, like for DWs or more originally relying on specific transport signature associated to the topological nature of skyrmions, e.g. NCMR, or topological Hall effect. Interestingly, this concept of skyrmion racetrack can be easily transformed and adapted to become a nanoscale voltage gate skyrmion transistor. This new function has been proposed by X. Zhang *et al.*[80] by adding a gate in a given part of track in order to locally modify though the application of an electric field, the magnetic properties of the magnetic media, being the perpendicular anisotropy or the DMI and thus controlling the passing or not of a skyrmion equivalent of the "on/off" switch of a transistor.

*Skyrmionic logic devices*

The fact that a skyrmion can be regarded as an independent 'particle' has also given rise to the proposal of several conceptual devices of skyrmion-based spin logic devices. Most of them rely on the results demonstrated through micromagnetic simulations[65] performed in nanoscale wires having different widths, that a single skyrmion can be transformed into a domain-wall pair and vice versa. This conversion mechanism allows in principle to duplicate or to merge skyrmions *at will* through the design of specific nanostructures and thus to perform basic logic operations. Based on these additional functionalities, X. Zhang *et al.*[81] have recently conceptualized skyrmion logic gates AND and OR, and thus realized the first step toward a complete logical architecture with the objectives to overtake the existing spin logic devices, particularly in their level of integration.

*Skyrmion magnonic crystal*

The ability to control spatially the nucleation of skyrmions, for example through either a local application of a magnetic[65] or electric field[67,82] or the injection of spin polarized current[41,42] give also the opportunity to prepare an artificial periodic arrangement of skyrmions in a 1D or 2D nanostructure. Such skyrmion lattices can then be used as periodic modulation of the magnetization to tailor the propagation of spin waves inside this novel type of "metamaterial". Indeed, F. Ma *et al.*[83] has recently shown through numerical simulations that a strong advantage of such skyrmion-based magnonic crystal compared to more standard ones (based on periodic modulation of the magnetic properties induced usually by lithography process) is that it can be dynamically reconfigured simply by changing the diameter of the skyrmions (by applying a magnetic field) or by changing the periodicity of the lattice or even erase it (Fig. 8c). Note also that skyrmion crystals at the scale of the nanometer can be envisaged, what is unreachable for conventional magnonic crystal fabricated with the existing lithography techniques. Leveraging of such functionality should thus allow a dynamical switching between full rejection and full transmission of spin waves in a waveguide. Besides the benefit for magnonic devices, it has been recently shown[84] that the spin waves themselves might be in a topological phase when propagating in a 2D atomic scale skyrmion lattice which should allow the realization of the spin-wave analogue of the anomalous quantum Hall effect for electrons.

*Skyrmion-based rf devices*

Another class of components for which the topological nature of skyrmions might induce a disruptive step is the nanoscale radiofrequency devices. For example, a dynamical mode of a single skyrmion in a dot that is typical of its topological nature is the low frequency breathing mode[85]. It has been proposed[86] that the skyrmion breathing mode induced by spin torques can be used to generate a rf signal if the dot with a skyrmion is part of a magnetoresistive device such as a spin valve or a magnetic tunnel junction (see Fig. 8d). One of the advantages of the resulting skyrmion-based spin torque oscillator, compared for example to a vortex based spin torque oscillator, can be that the skyrmion, being a localized soliton, will be less sensitive to external perturbation and thus display a more coherent dynamic. Another function numerically investigated by G. Finocchio *et al.*[87], is the concept of skyrmion-based microwave detector, which relies on the resonant excitation of the breathing mode when the frequency of the external rf signal equals the frequency of the breathing mode (that can be largely changed by the application of an external perpendicular field for example) and the conversion of this resonant dynamics into a dc mixed voltage. Finally, a novel type of skyrmion-based spin torque oscillator recently proposed by F. Garcia-Sanchez *et al.*[88] is based on the self-sustained gyration arising from the competition between the confinement from the boundary edges and the spin forces due to an inhomogeneous spin polarizer. The associated gyrotropic frequency is lower by about one order of magnitude compared to conventional vortex based spin torque oscillators but their main advantage is that there is no threshold current for the onset of the skyrmion dynamics.

**Outlook**

The seminal observations of magnetic skyrmions in 2009 have kicked off an intense research effort that has rapidly led to significant advances on both the understanding of the physical mechanisms and the realization of suitable structures for applications. It is now possible to stabilize small individual compact skyrmions at room temperature in multilayers fabricated by usual technologies of spintronic devices. Their current-induced motion has been clearly observed with velocities up to the 100 m/s range and the mechanisms of the motion begin to be understood. In addition, several concepts for a controlled creation and detection of skyrmions have been demonstrated. Some pending questions have still to be solved. For example, a better control of the size of the skyrmions in multilayers is necessary to obtain the most convenient sizes for applications (probably in the 10 nm range). A better understanding of the role of defects and influence of homogeneities of the material is needed as well, to achieve an uniform motion of assemblies of skyrmions. However, the recent rapid advances of the research about the skyrmions make us optimistic for the realizations at medium term of some of the devices that have been already modelled. Besides the ones described, many other concepts of device can also be anticipated, as, for example, specific components for bio-inspired computing[89,90] or multi-state memories based on the geometrical configuration of an ensemble of skyrmions.

**BOX 1: Skyrmion lattices and metastable individual skyrmions**

In ultra-thin films submitted to strong DMI, Fe monolayer[10] or Fe/Pd bilayer[11] on Ir(111) for example, the ground state at zero field is a non-collinear spin texture such as a spin spiral or a skyrmion lattice, and, generally, a large magnetic field is necessary to induce a ferromagnetic state embedding metastable individual skyrmions, see Fig. 1d for Fe/Pd bilayer in 1.5 T. A transition from skyrmion lattice to ferromagnetic ground state occurs by applying a field or, at zero field, when $D$ becomes smaller than a critical value $D_C = 4(AK)^{1/2}/\pi$ in an analytical treatment of a continuous magnetization model[91] ($A$ and $K$ are exchange stiffness and effective out-of-plane anisotropy respectively). Actually, due to magnetostatic interactions, the zero field ground state for $D < D_C$ is not always the saturated ferromagnetic state but can be a state with magnetic domains. The domains collapse by application of a generally small field and a ferromagnetic phase including topologically protected metastable skyrmions can be obtained in an intermediate field range (typically between 10 and 100 mT) before complete saturation a higher field, see Fig. 3b with successively domains, skyrmions and saturation at increasing field in Ir/Co/Pt multilayers[31]. Several other ways to transform the phase with magnetic domains into a phase with metastable individual skyrmions, by field or current pulses, will be discussed in other parts of the article.

The Figure from Siemens *et al.* (adapted with permission from REF.[92]), derived from Monte Carlo simulations with discrete atomic spins, displays the succession of different phases as $D$ decreases at constant exchange, anisotropy, applied field and temperature: spin spiral and skyrmion lattice for large $D$ on the right, saturated ferromagnetic state on the left and, between the dashed lines, ferromagnetic ground state embedding individual skyrmions. The transition from lattice to individual skyrmion, with a smooth maximum of the skyrmion radius at the transition, is much less abrupt than expected by analytical continuous models of infinite samples in which the skyrmion radius tends to infinity at the transition[91]. This likely comes from simulations with finite sample size at finite temperature. The collapse of skyrmions and saturation depends also on temperature which controls the escape over the topological barrier. At zero temperature, the transition to saturated state from the skyrmion phase would be shifted to the left and occur at sizes of the order of the atomic distance.

The transition between skyrmion lattice and individual skyrmion in experiments of Fig. 3d is also far from a sharp transition. By varying the relative thicknesses of Fe and Co in the magnetic layers to shift the ratio $D/D_C$ (labelled as $\mathcal{K}_{est}$ in Fig. 3d) from $D/D_C < 1$ (in the grey zone in the left of the graph) to $D/D_C > 1$, there is a progressive transition from disordered individual skyrmions (bottom image) to hexagonal skyrmion lattice (top image) with partly ordered states in between. Such a progressive transition, likely due to inhomogeneity, defects and finite size, is an interesting phenomenon by itself but it certainly complicates the quantitative analysis of experiments.

**BOX 2: Skyrmions and chiral bubbles**

Several recent experiments have found "big skyrmions", with diameter in the µm range, that have been described as chiral or skyrmionic bubbles[35], that is domains of reversed magnetization surrounded by a Néel domain wall of the chirality favoured by the DMI. These chiral bubbles have a topological number equal to unity as a prototypic "compact" skyrmion[3] and are different from classical bubbles without any preferred chirality stabilized solely by dipole interactions. Chiral bubbles and standard skyrmions, in spite of their large difference in size but thanks to their topological similarity, share several properties as for example similar current-induced motions. Kiselev *et al.*[93] have shown that DMI can lead to the coexistence of skyrmion and bubble solutions in a rather narrow range of the material parameters. The scheme of the top figure [adapted with permission from REF.93] shows that, by increasing the DMI from zero where there exists a single energy minimum (for radius $R_1$, black curve) corresponding to a classical bubble to (in red) three minima corresponding respectively to a skyrmion (at $R_S$), a chiral bubble with chirality favoured by DMI (at $R_2$) or in opposition to DMI (at $R'_2$) with a larger energy. The bottom figure shows the difference between the profile of a compact skyrmion (its diameter essentially corresponds to the DW width) and that of a chiral bubble with a large plateau of magnetization in the centre and a sharp reversal at the scale of a DW width at the approach of the edge. The question of a possible continuous transition from skyrmions to chiral bubbles has not yet been really investigated and quantitatively described.

**Acknowledgements:**

The authors acknowledge K. Bouzehouane, S. Collin, K. Garcia, W. Legrand, D. Maccariello, C. Moreau-Luchaire for their participation to on-going studies on skyrmions and C. Panagopoulos, S. Rohart, J. Sampaio, A. Thiaville as well as all the partners involved in MAGicSky consortium for fruitful discussions. Financial support from European Union grant MAGicSky No. FET-Open-665095. 103 is acknowledged.

**Figure captions:**

**Fig. 1: Magnetic texture of skyrmions.** (**a-b**) Néel-like (**a**) and Bloch-like (**b**) skyrmions with magnetization rotation from the skyrmion centre to the external uniform magnetization as in a Néel domain wall (a) or a Bloch one (b). Reproduced with permission from REF. 94. (**c**) Lattice of skyrmions as observed by spin-polarized scanning tunneling microscopy in a monolayer of Fe grown on top of Ir(111). Adapted with permission from REF.95. The color wheel indicates the in-plane magnetization. (**d**) Individual skyrmions observed by the same technique in PdFe|Ir(111). The out-of-plane magnetization is color-coded from red for 'up' to blue for 'down' magnetization. An external field B = 1.5 T is used to stabilize these skyrmions. Adapted with permission from REF.96.

**Fig. 2: Dzyaloshinskii-Moriya interactions (DMI).** (**a**) Simplest picture[5,13] of interfacial DMI mediated by electrons interacting by exchange with magnetic spins and spin-orbit coupled on heavy metal sites at interfaces between magnetic and heavy metals, here Co and Pt. Adapted with permission from REF.5 (**b**) Ab-initio calculation of DMI between pairs of atomic spins in monolayers of 3d metals on 5d metals. Adapted with permission from REF.19. (**c,d**) Distribution of DMI between pairs of atomic spins in the different magnetic layers of (Pt-3ML/Co-3ML) (c) (Adapted with permission from REF.18) and (Pt-3ML/Co-5ML/MgO) (d) (Adapted with permission from REF.97.) from ab-initio calculations. The predominant DMI are at the Interface of Co with Pt but significant DMI also exist at the interface with MgO.[19]

**Fig. 3: Experimental observation of skyrmions in magnetic multilayers with different imaging techniques.** (**a**) Illustration of the additive DMI induced by different heavy metal (here Ir and Pt) sandwiching a magnetic layer. Adapted with permission from REF.31. (**b-c**) Sub-100nm skyrmions observed by STXM in [Pt|Co|Ir] multilayers and compressed down to less than 20 nm radius by an out-of-plane field. (**b**) As the field is decreased, the system goes from saturation (top right image, 83 mT) to a labyrinth-shaped domain structure (top left, 8 mT), passing through states containing skyrmions in the 10-70 mT range. In the profiles of the out of plane magnetization at 38 and 68 mT, the black dots are from STXM measurements, the blue dotted lines are theoretical skyrmion profiles, and the red lines are fits with the experimental profiles obtained after convolution with the STXM beam shape. Reproduced with permission from REF.31. (**c**) The radius of the skyrmions is plotted against the out-of-plane magnetic field (black dots) and compared with micromagnetic simulations for different magnitudes of the DMI. Adapted with permission from REF.31. (**d**) Magnetic force microscopy (MFM) images of skyrmions for different Fe/Co ratios and different values of $\kappa = D/D_c$ in [Pt|Co|Fe|Ir] multilayers. A smooth transition from a state with disordered individual skyrmions (bottom image) to skyrmion lattice (top image) as $\kappa = D/D_c$ increases is observed. Reproduced with permission from REF.32. (**e**) XMCD-PEEM image of a skyrmion in a 420 nm square (dotted line) in [Pt|Co|MgO] Adapted with permission from REF.20. (**f**) SPLEEM images of skyrmions in the [Fe|Ni] layers of the structure shown in (g). (**g**) Schematic of the multilayer in which skyrmions in Fe/Ni displayed in (f) are stabilized in zero external field through exchange with a Ni layer. Panels f and g adapted with permission from REF.36.

**Fig. 4: Current-induced motion of skyrmions.** (**a**) Micromagnetic simulations of motion by STT: skyrmion velocity in 0.4 nm thick Co (α = 0.3, see text) on heavy metal layer vs current density with STT due to SHE in heavy metal of typical spin Hall angle equal to 0.1 (blue squares) and (red dots) for the STT due to a 40% spin polarized current in Co with, for the non-adiabatic torque coefficient, β = α = 0.3 The STT from SHE is one order of magnitude more efficient than the conventional STT in the magnetic layer. (Adapted with permission from REF.41.) The insets illustrate both configurations: For SHE STT the spin current propagates from the heavy metal layer (greenish colour), while for CIP STT the spin currents is a spin polarized charge current along the magnetic track. (**b**) Illustration of the skyrmion motion in a track: the motion starts with a transverse component, bends due to the repulsion by the edge and then propagates along the edge. Reproduced with permission from REF.41. (**c-d**) Estimation of the velocities of skyrmions and chiral bubbles, from compact skyrmions ($R/\Delta = \pi/2$) to large bubbles, as calculated from Thiele equations under a driving force due to spin Hall effect and for two values of the damping parameter α. The skyrmion Hall angle (c) decreases as the size of the texture increases, while the velocity normalized by the spin current $J_s$ (d) first increases with the size of the core of the chiral bubble and then saturates ($v_x$) or decreases ($v_y$).

**Fig. 5: Observations of skyrmion motions.** (**a**) Skyrmion Hall Effect, characterized by a transverse velocity component, as observed by Kerr microscopy in Ta|CoFeB|TaO$_x$ trilayer ($j_e$ is the electronic current density). Inset: velocity versus current density. Adapted with permission from REF.47. (**b**) Velocity versus current density for skyrmions in [Pt|Co|Ta] and [Pt|CoFeB|MgO] multilayers observed by STXM. Reproduced with permission from REF.33. In the inset of (a) as well as in (b), the velocity takes a finite value only above a critical current (~2.5 $10^{11}$ A/m$^2$) and then increases almost linearly, see discussion in text. (**c**) Velocity vs current density in (Pt|Ni|Co|Ni|Au|Ni|Co|Ni|Pt) multilayer for coupled skyrmions of opposite chiralities in the two Ni|Co|Ni trilayers (see inset for the mechanism of magnetostatic coupling) and displaced in the same direction by the SHE of the top and bottom Pt layers. Reproduced with permission from REF.48.

**Fig. 6: Techniques for skyrmion nucleation.** (**a**) Snapshots of dynamical spin configurations around a notch during the process of skyrmion creation by the action of the current-induced STT on the in-plane spin components near the boundary (micromagnetic simulations). Reproduced with permission from REF.42. (**b**) Device for vertical spin injection through a magnetic tunnel junction and micromagnetic simulations showing a final state with skyrmion after a 2ns current pulse of 3x$10^{11}$ A/m$^2$ and spin polarization of 40%. Adapted with permission from REF.41. (**c**) Example of nucleation of skyrmionic bubbles in Ta|CoFeB|TaO$_x$ by the current-induced expansion of domains at the exit of a constriction. Adapted with permission from REF.35. (**d**) Snapshot in the progressive curving of a domain wall during micromagnetic simulations of the SHE-induced creation of skyrmions at the exit of a constriction. Reproduced with permission from REF.63. (**e**) Deleting (top images) or writing (bottom images) skyrmions with an electric field (schematic of experiments on the left) in 3 ML of Fe on Ir(111).The skyrmion distortion reflects the strain of the lattice. Adapted with permission from REF.67.

**Fig. 7: Techniques for skyrmion detection.** (**a**) Illustration of a skyrmion racetrack with MTJ for skyrmion readout. (**b**) NCMR: differential tunnel conductance spectrum (d$I$/d$U$ vs voltage $V$) with a nonmagnetic STM tip at the centre of a skyrmion (red curve) and outside the skyrmion (black curve) in PdFe/Ir(111). Reproduced with permission from REF.69. (**c**) Topological Hall effect (THE) signal as a function of the distance $x$ between a skyrmion and the voltage contacts, as calculated by Hamamoto *et al*. (adapted with permission from REF.71), and (**d**) experimentally observed in an actual device with track width of the same order of magnitude than the skyrmion size. Adapted with permission from REF.74.

**Fig. 8: Skyrmions for applications.** (**a**) Train of skyrmions in a racetrack. (**b**) Micromagnetic simulations showing that skyrmions are compressed in a track, effectively reducing their size to adjust to the track width, and hence increasing the information density. Note that the skyrmions keep a circular shape. (**c**) A skyrmion lattice could form an ideal magnonic crystal with extremely small dimensions, unreachable with conventional lithography. Adapted with permission from REF.83.(**d**) Different excitation modes of the skyrmions could be used to fabricate radiofrequency filters. Adapted with permission from REF.86.

[97] Yang, H., Boulle, O., Cros, V., Fert, A., & Chshiev, M., Controlling Dzyaloshinskii-Moriya Interaction via Chirality Dependent Atomic-Layer Stacking, Insulator Capping and Electric Field. arXiv: 1603.01847v2 (2016).

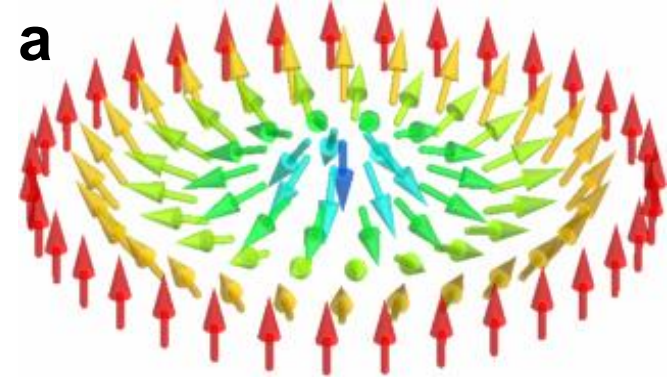
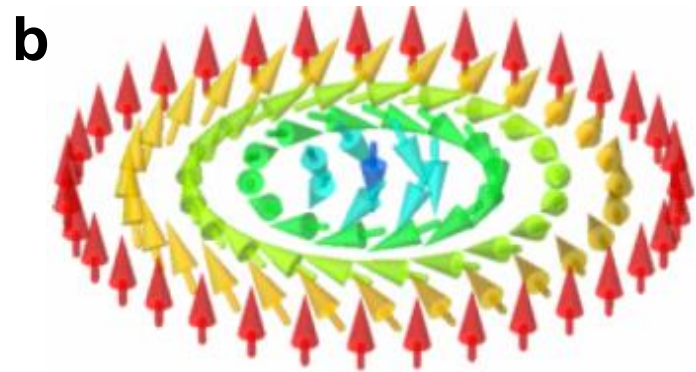
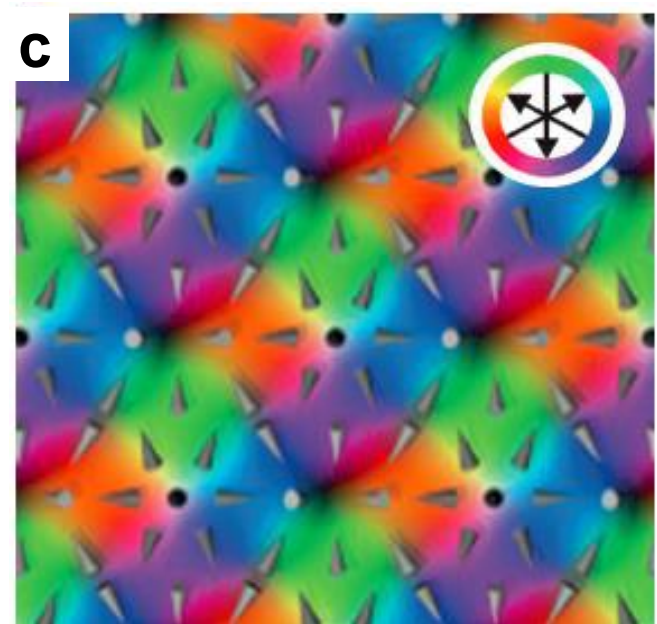
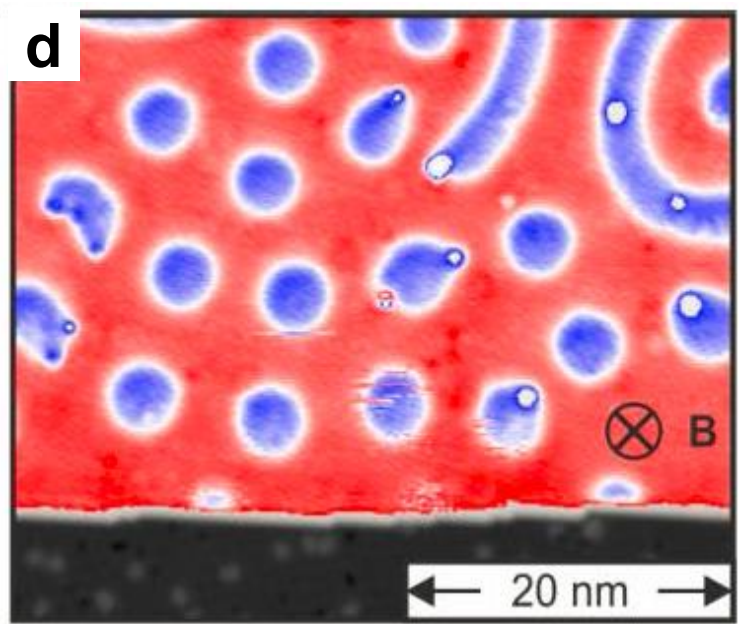

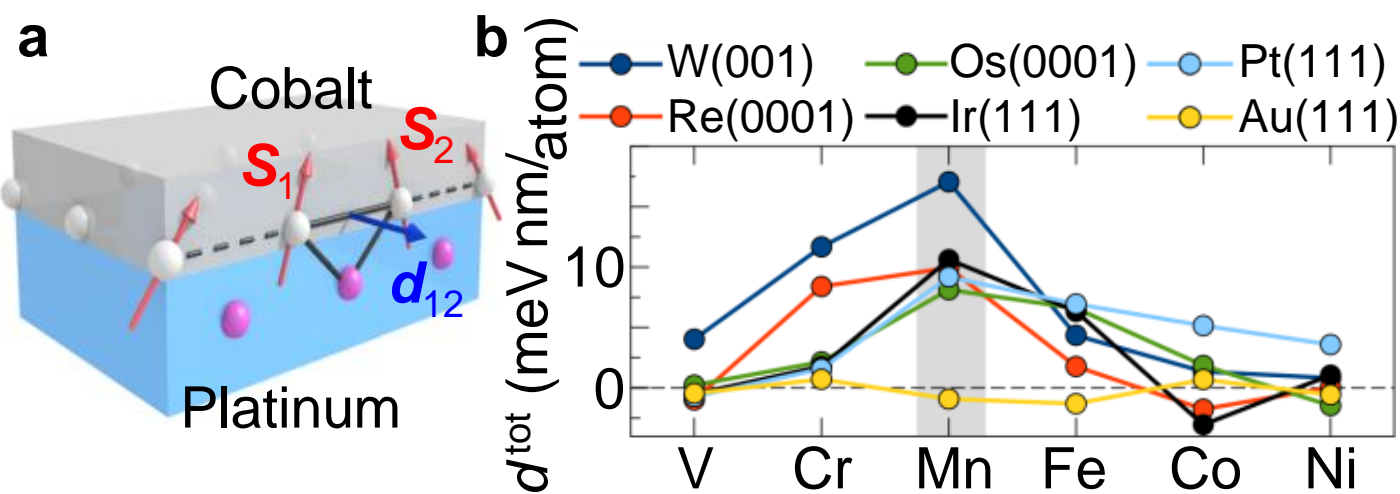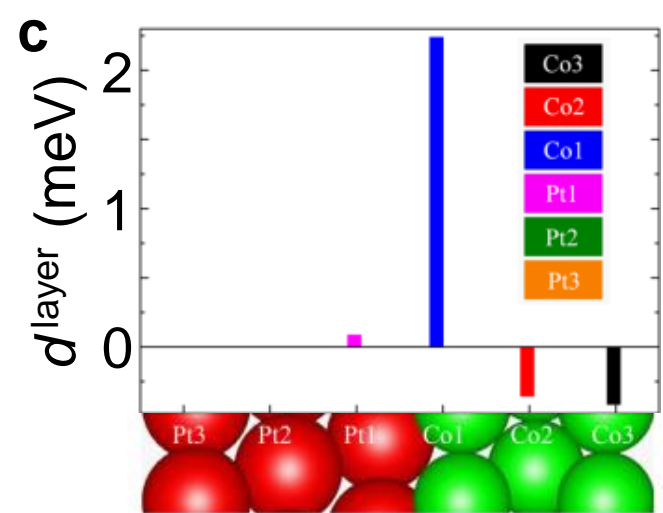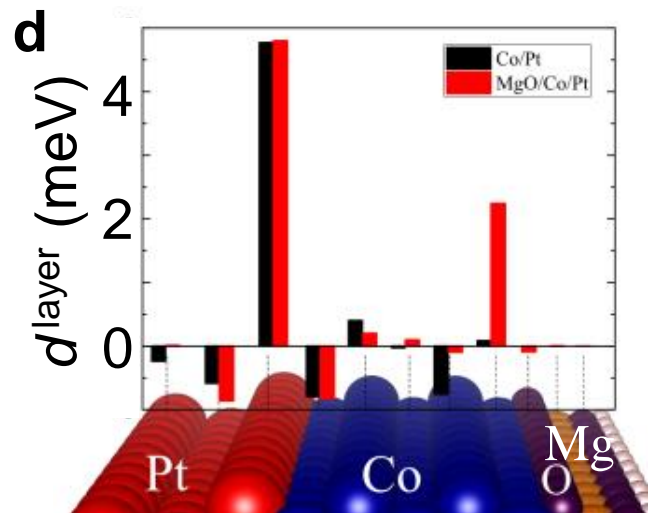

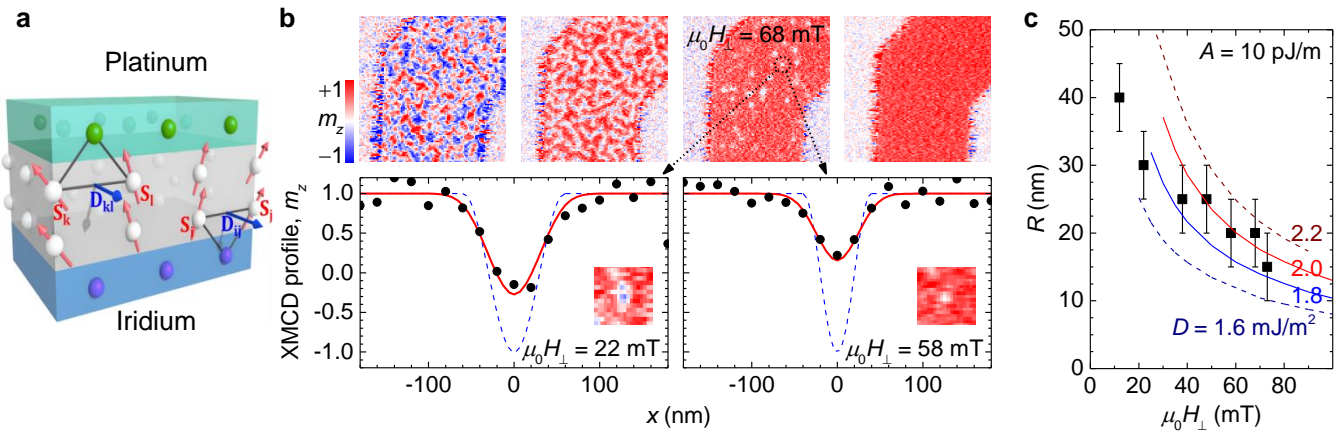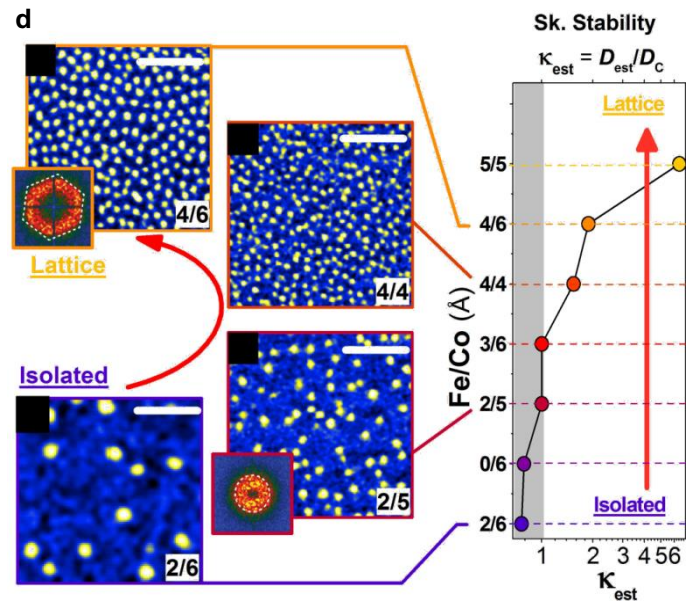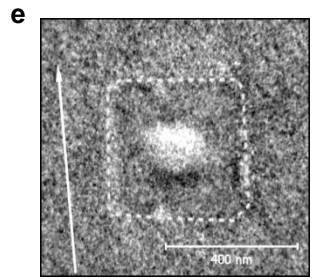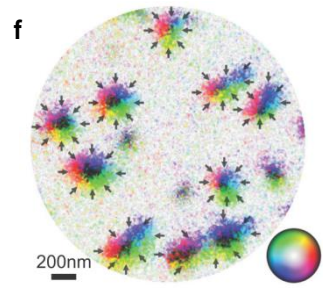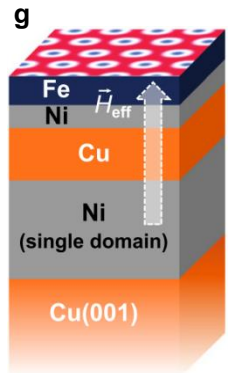

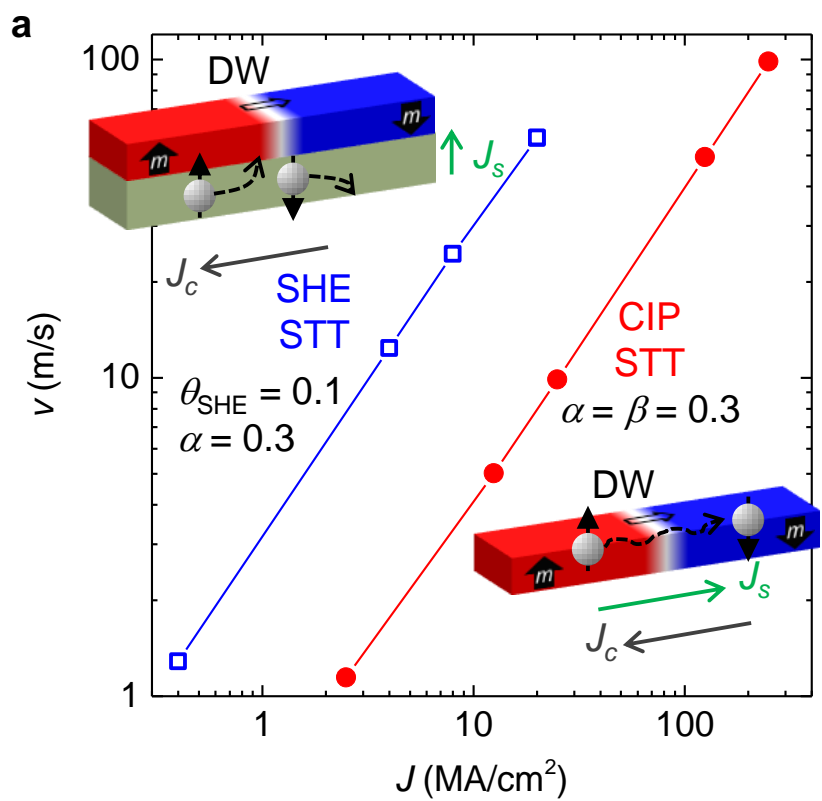
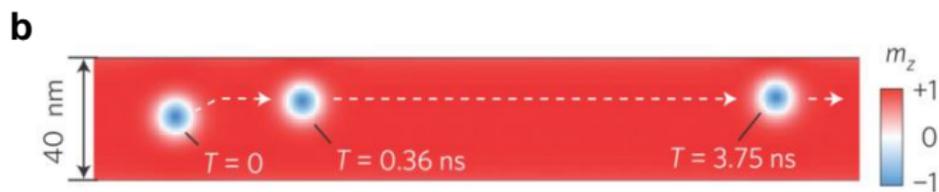
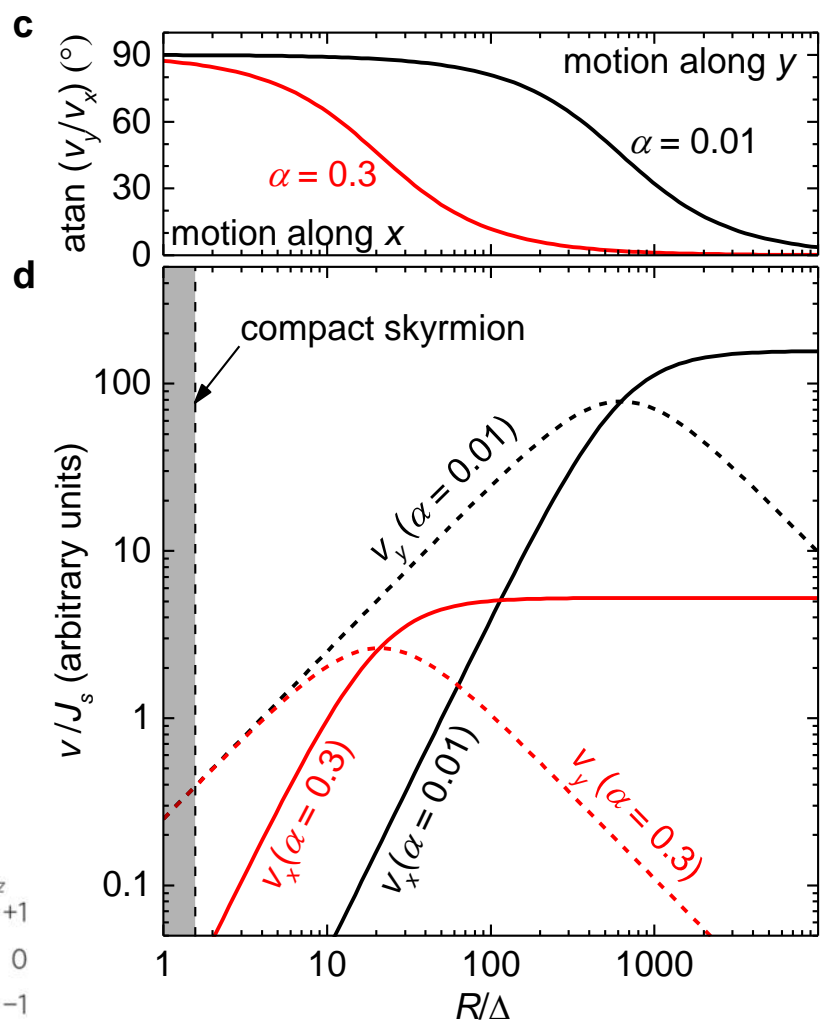

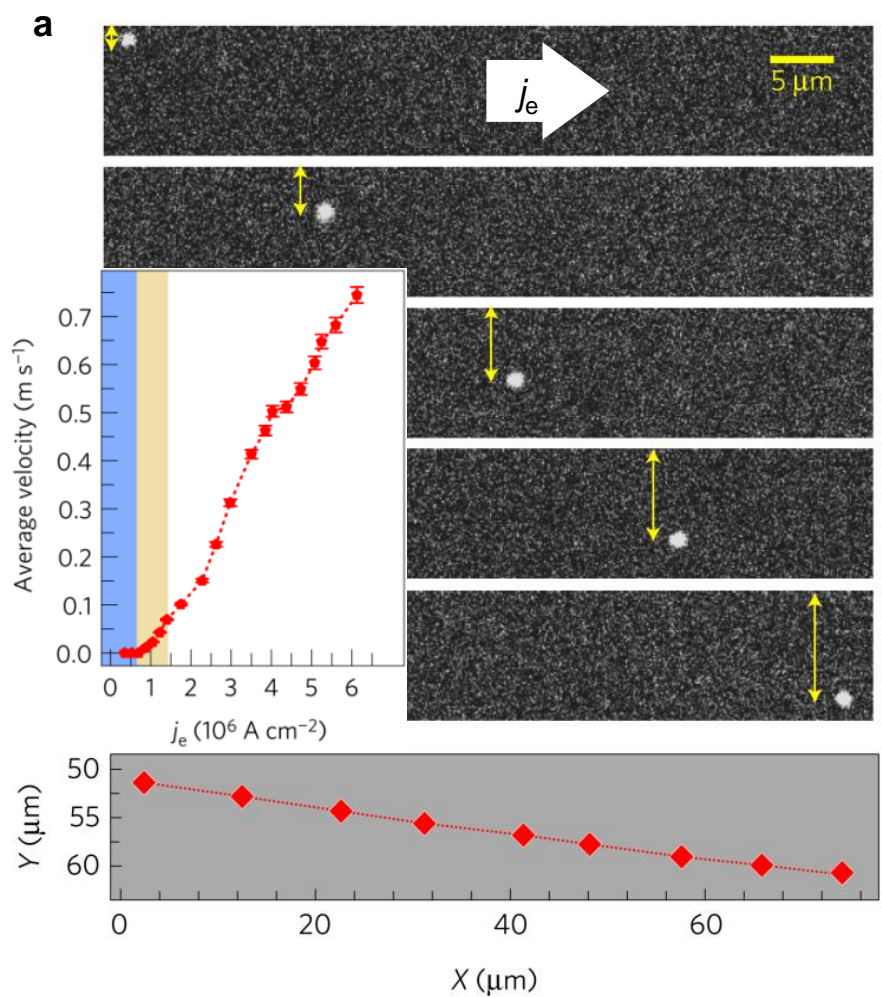
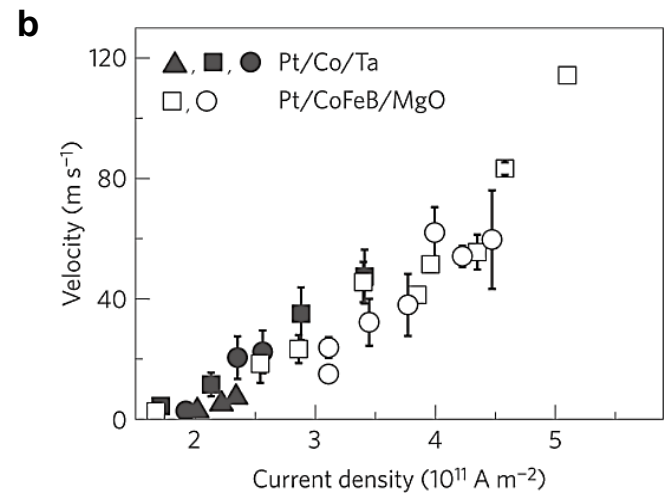
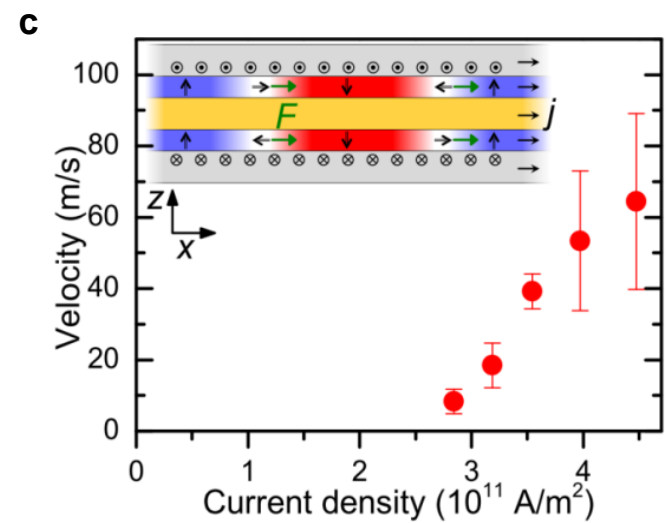

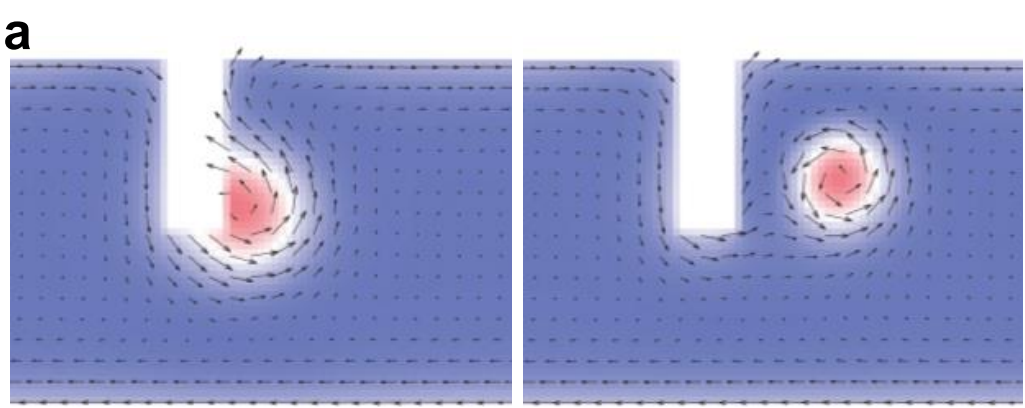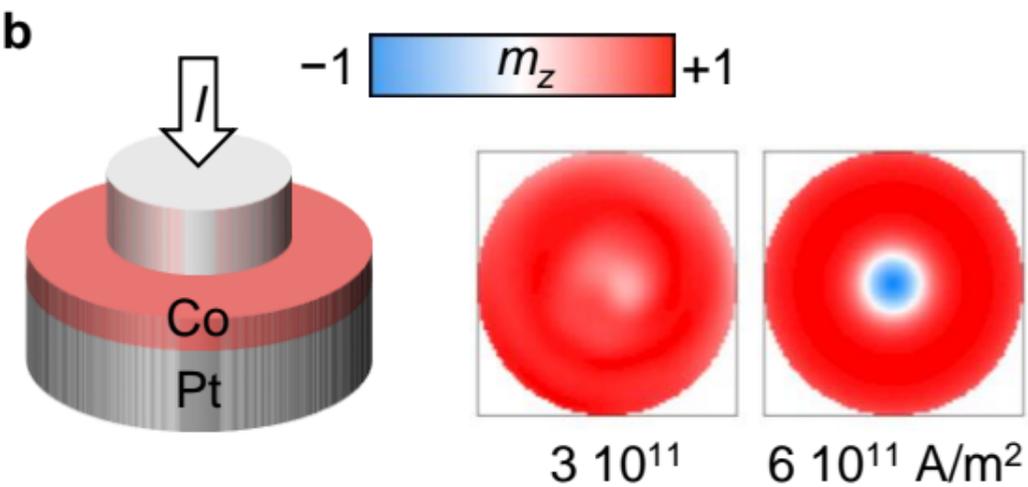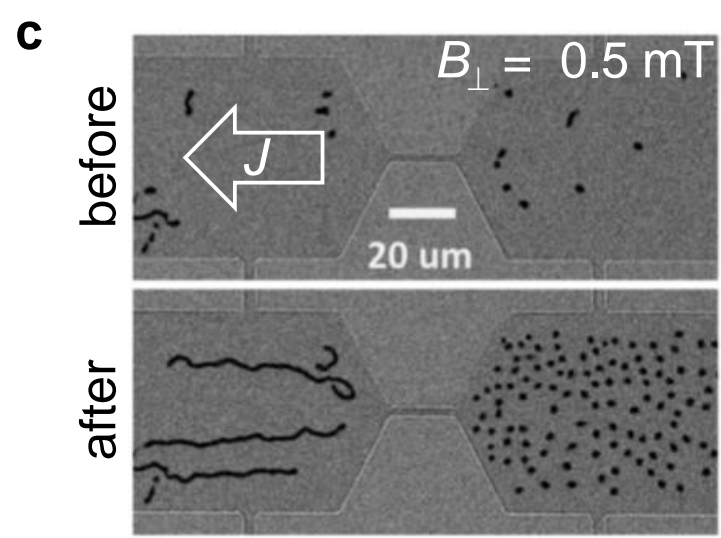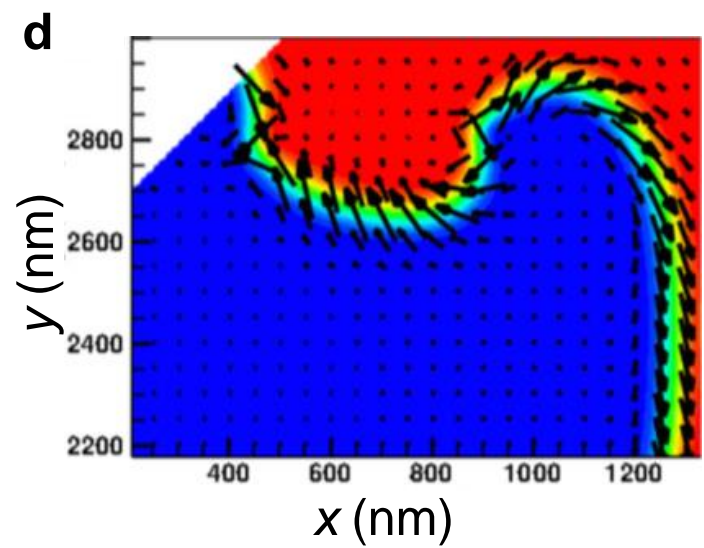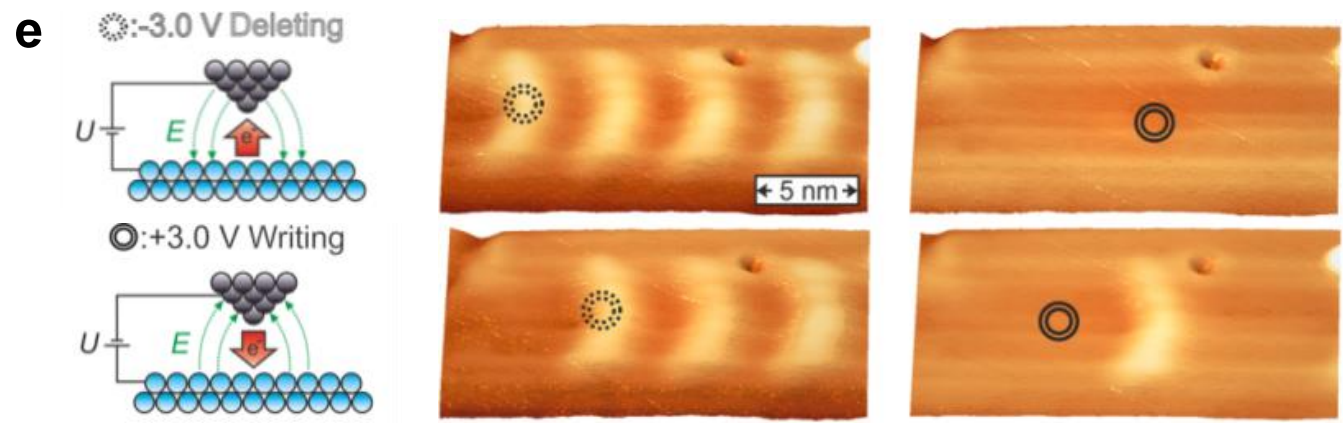

**a** 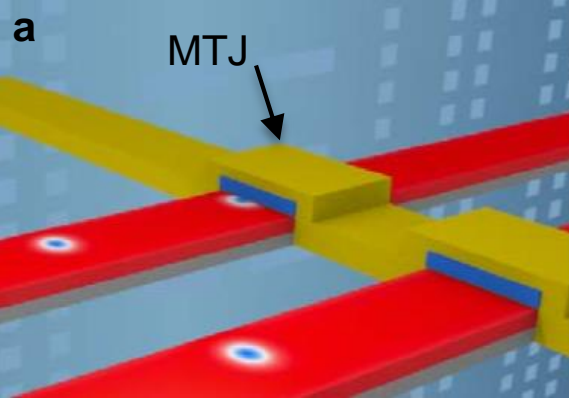

**b** 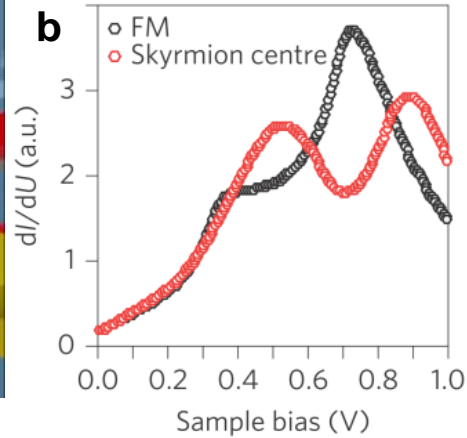

**c** 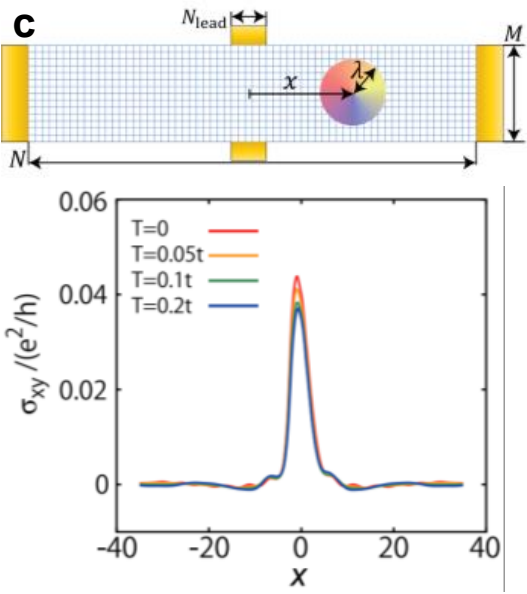

**d** 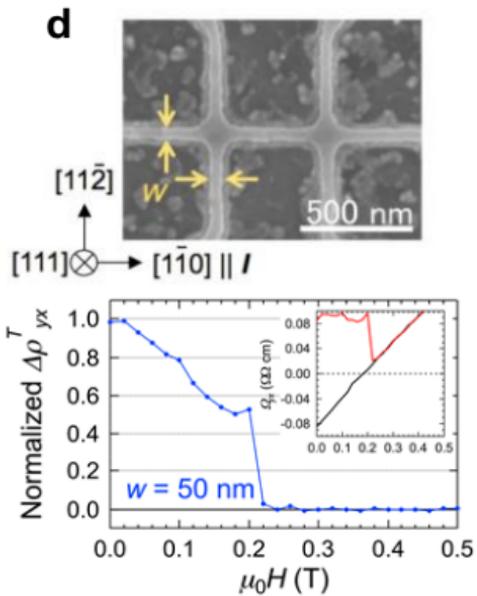

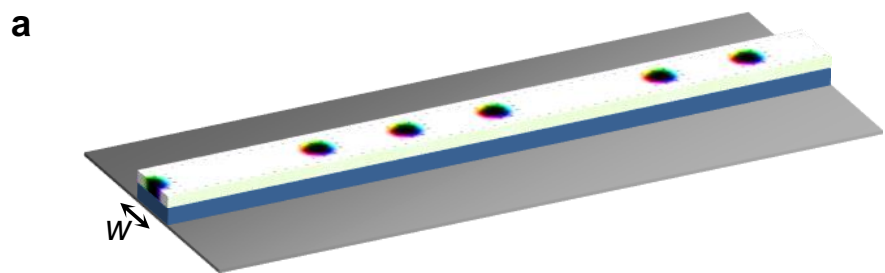

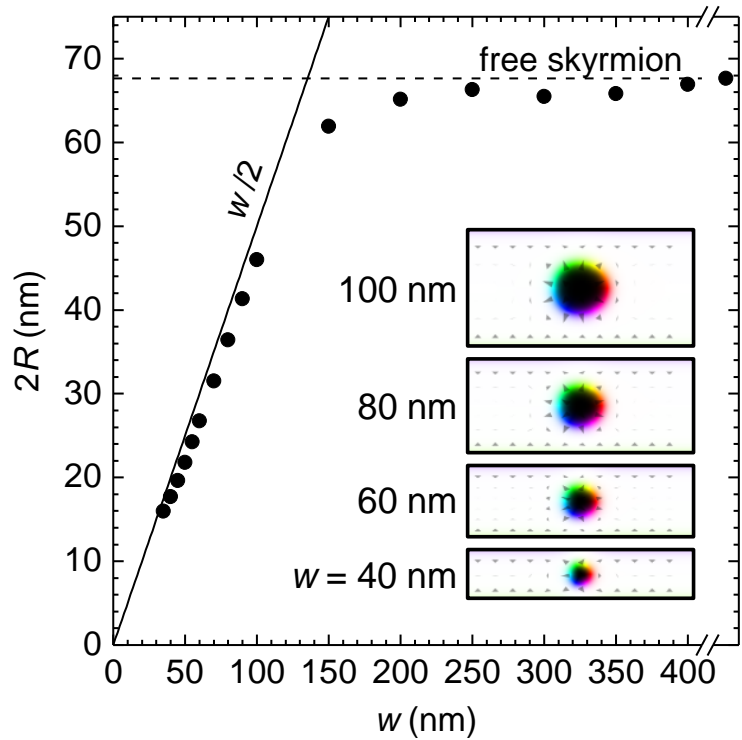

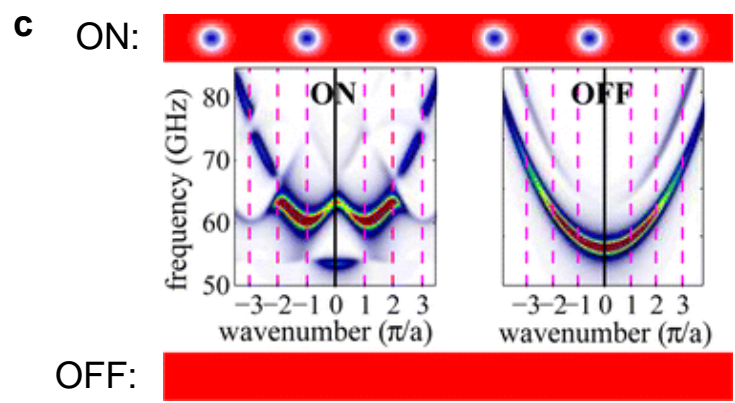

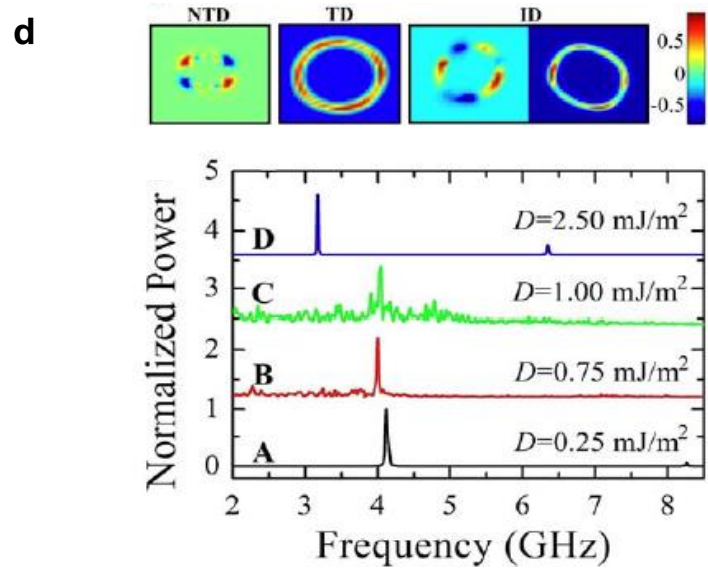

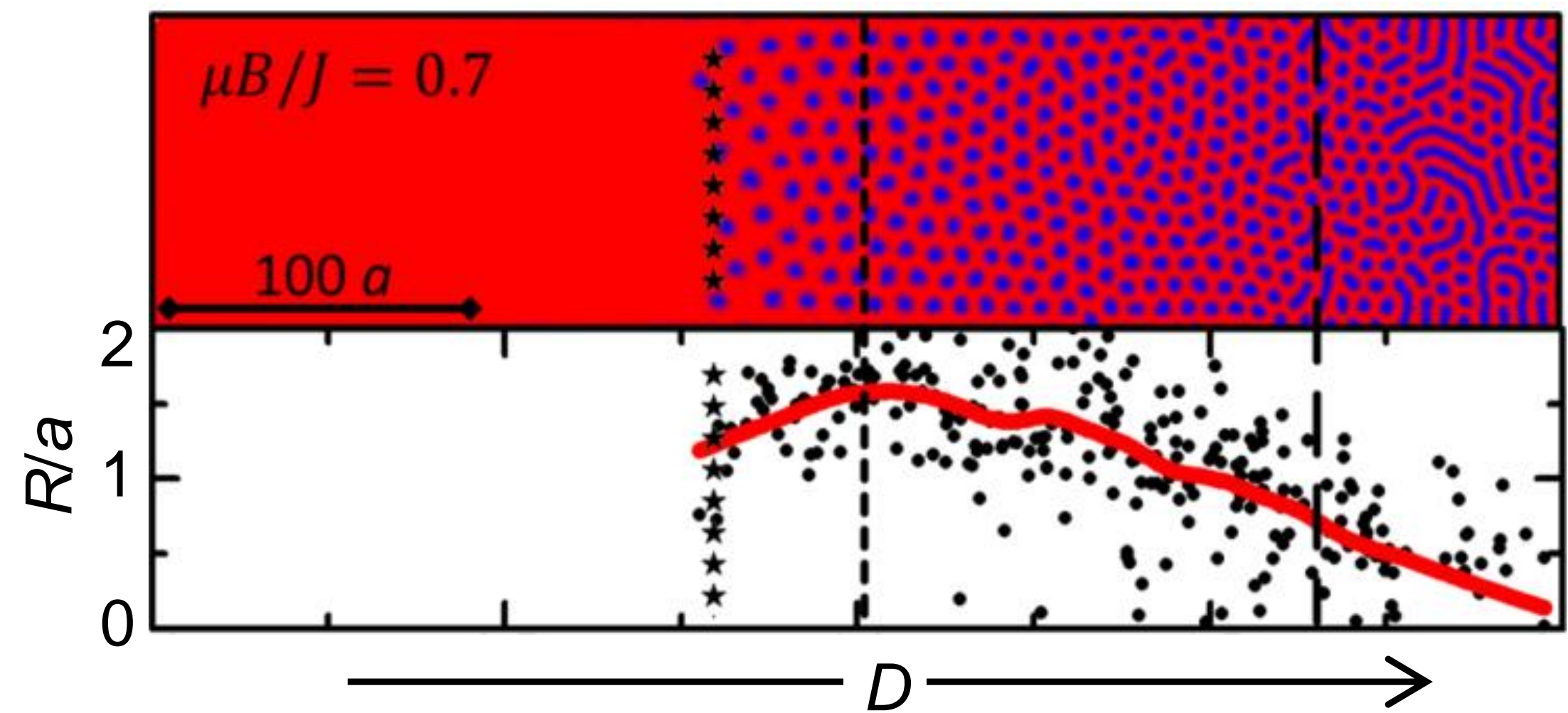

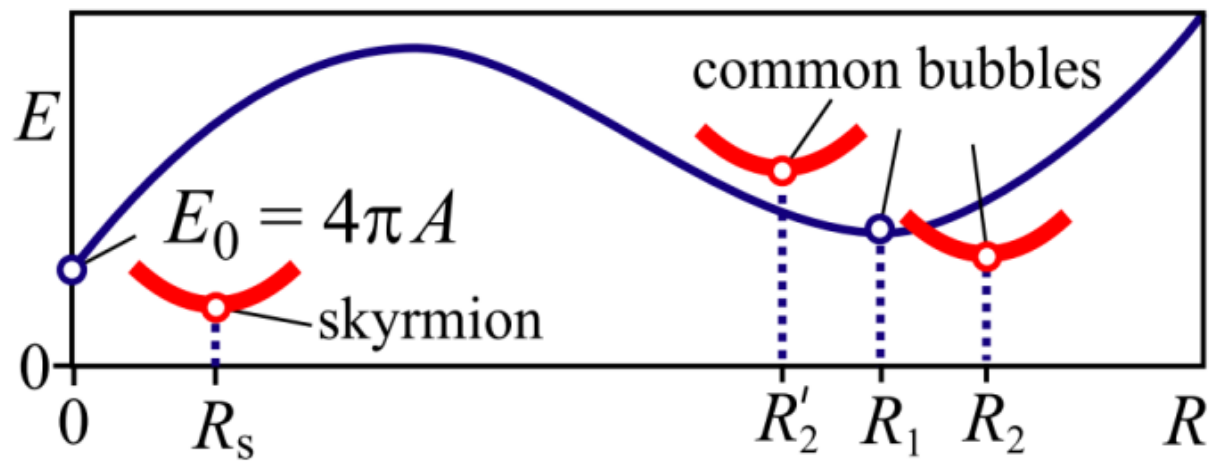
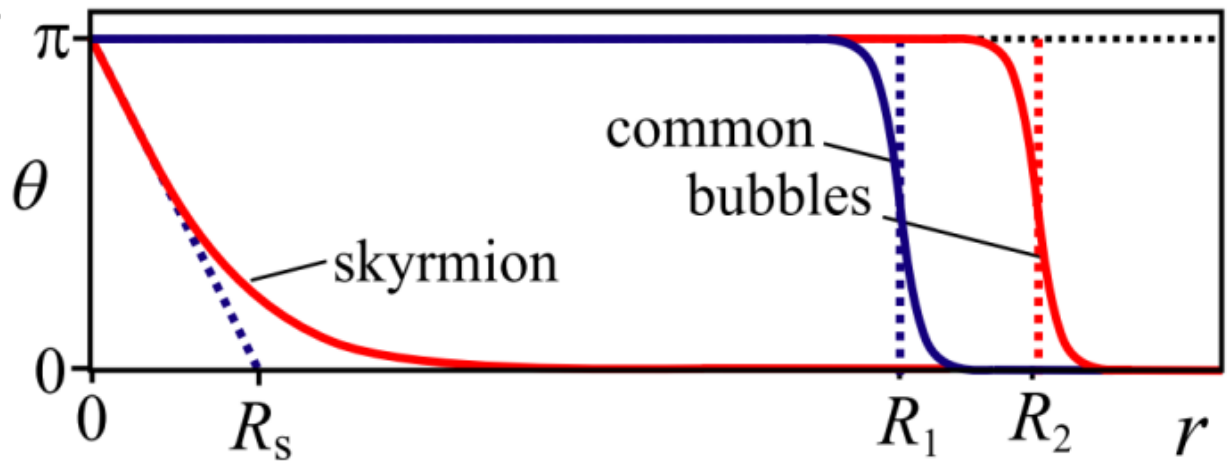